\documentclass[twocolumn]{webofc}
\usepackage[varg]{txfonts}   
\begin{document}
\title{Effect of Pauli repulsion and transfer on fusion}
%
%

\author{\firstname{C.} \lastname{Simenel}\inst{1}\fnsep\thanks{\email{cedric.simenel@anu.edu.au}} \and
        \firstname{K.} \lastname{Godbey}\inst{2} \and
        \firstname{A. S.} \lastname{Umar}\inst{2} \and
        \firstname{K.} \lastname{Vo-Phuoc}\inst{1} \and
        \firstname{M.} \lastname{Dasgupta}\inst{1} \and
        \firstname{D. J.} \lastname{Hinde}\inst{1} \and
        \firstname{E. C.} \lastname{Simpson}\inst{1} 
}

\institute{Department of Nuclear Physics, Research School of Physics and Engineering, The Australian National University, Canberra ACT  2601, Australia
\and
           Department of Physics and Astronomy, Vanderbilt University, Nashville, TN 37235, USA
           }

\abstract{%
The  effect of the Pauli exclusion principle on the nucleus-nucleus bare potential is studied using a new density-constrained extension of the Frozen-Hartree-Fock (DCFHF) technique. 
The resulting potentials exhibit a  repulsion at short distance. The charge product dependence of this Pauli repulsion is investigated. 
Dynamical effects are then included in the potential with the density-constrained time-dependent Hartree-Fock (DCTDHF) method.
In particular, isovector contributions to this potential are used to investigate the role of transfer on fusion, resulting in a lowering of the inner part of the potential for systems with positive Q-value transfer channels. }
\maketitle
\section{Introduction}
\label{intro}
A dream shared by many theorists working on the quantum many-body problem is to find a way to describe the tunnelling of a many-body wave-function.
For instance, this would enable a fully microscopic description of sub-barrier fusion, without other parameters than those of the energy density functional describing the interaction between the nucleons. 

Such a tool could then be used to investigate the longstanding deep sub-barrier fusion hindrance puzzle~\cite{jiang2002,dasgupta2007,stefanini2010}   (see Ref.~\cite{back2014} for a review).
It is also crucial to predict fusion cross-sections in systems such as $^{12}$C$+^{12}$C at astrophysical energies in order to 
get a deeper insight into stellar nucleosynthesis mechanisms  (see, e.g., Ref. \cite{santiago-gonzalez2016} and the contributions to this conference from S.~Courtin, G.~Fruet, E.~Rehm, and N.~T.~Zhang for recent works). 
In addition, it would help guiding experimental programs aiming at studying the impact of exotic structures (e.g., neutron-skins and pigmy dipole resonances) and of the continuum on fusion (see Refs. \cite{desouza2013,kolata2016,singh2017} and contributions from D.~Bazin, J.~Kolata, R.~T.~de~Souza and G.~Colucci for new or recent experimental programs). 

However,  such a theory is not yet available. 
Indeed, microscopic descriptions of fusion reactions are based on mean-field approximations, such as the time-dependent Hartree-Fock (TDHF) theory, which do not account for tunnelling of the many-body wave function.
This is because more than one mean-field is required to describe the outcome of a near-barrier reaction (one for the fused system and one for the outgoing fragments). 

\begin{figure}
\centering
\includegraphics[width=4cm]{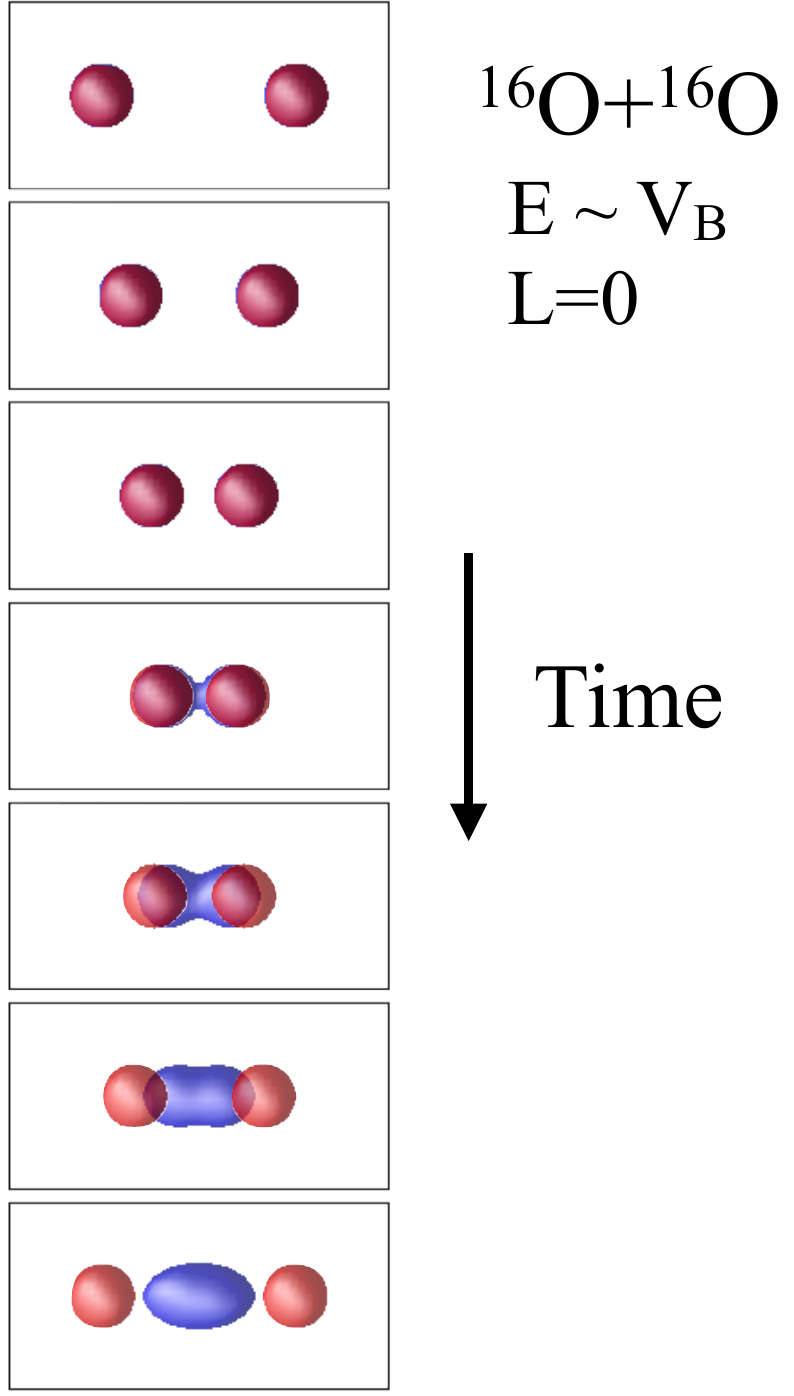}
\caption{Example of isodensities obtained from TDHF calculations of $^{16}$O$+^{16}$O central collisions at energies slightly below (red) and just above (blue) the fusion barrier. Adapted from~\cite{simenel2014b}. }
\label{fig:tunnelling}       
\end{figure}

Figure~\ref{fig:tunnelling} shows TDHF density evolutions for $^{16}$O$+^{16}$O central collisions at near barrier energies, one leading to fusion (just above the barrier) and one leading to two outgoing fragments (below the barrier). 
We see that both collisions lead to density distributions occupying different regions of space. 
Thus, sub-barrier fusion, associated with non-zero fusion and scattering probabilities, cannot be described with a single local mean-field. In TDHF, fusion probabilities are then either 0 or 1 for a given initial condition. 

A (temporary) solution is to reduce the many-body description to a two-body system (though it often oversimplifies the problem) where the fusion probability is computed from the transmission through a microscopically derived nucleus-nucleus potential (see, e.g., Refs.~\cite{umar2006a,simenel2013a,simenel2013b,umar2014a}). 

In the present contribution we review recent microscopic methods to determine the nucleus-nucleus potential and use them to predict fusion cross-sections. First, we discuss the standard frozen Hartree-Fock (FHF) approach in section~\ref{sec:FHF}. We then extend this approach in section~\ref{sec:Pauli} to  incorporate the  effect of the Pauli exclusion principle between nucleons belonging to different collision partners using a new density-constrained extension of the Frozen-Hartree-Fock (DCFHF) technique. The effect of Pauli repulsion on deep sub-barrier fusion is then discussed in section~\ref{sec:subB}.
Finally, dynamical effects are  studied in section~\ref{sec:dynamics} using the density-constrained time-dependent Hartree-Fock (DCTDHF) method. In particular, the isovector contributions to the dynamical potential are used to get a deeper insight into the effect of transfer on fusion. 

\section{The frozen Hartree-Fock (FHF) method}\label{sec:FHF}

A standard approach is to compute the bare nucleus-nucleus potential from frozen static Hartree-Fock (HF) ground-state densities \cite{simenel2008,washiyama2008} using the energy density functional approach of Brueckner {\it et al.} \cite{brueckner1968}.
The HF ground-state of a nucleus is obtained by solving the variational principle 
\begin{equation}
\delta\langle\Phi|\hat{H}|\Phi\rangle =0, 
\label{eq:HF}
\end{equation}
where $|\Phi\rangle$ is an independent many-fermion state which can be written as  a Slater determinant of the occupied single-particle states, ensuring a full account of the Pauli exclusion principle in the ground-state density.

At the mean-field level, the state $|\Phi\rangle$ contains the same information on the system as the one-body density matrix $\rho$. 
The energy of the system can then be written as an energy density functional (EDF) $E[\rho]$. 
In practice, the Skyrme EDF \cite{skyrme1956} is often used, where the energy density $\mathcal{H}(\mathbf{r})$ only depends on the local part of $\rho$ and of its derivatives,
\begin{equation}
E[\rho]=\int d^3r\,\,\, \mathcal[\rho(\mathbf{r}),\tau(\mathbf{r}), \mathbf{J}(\mathbf{r})\cdots],
\end{equation}
where $\rho(\mathbf{r})$, $\tau(\mathbf{r})$, and $\mathbf{J}(\mathbf{r})$ are the local particle, kinetic, and spin-orbit densities, respectively. 
These densities are symmetric under time-reversal transformations. 
For systems for which this symmetry does not hold, other densities such as the current density $\mathbf{j}(\mathbf{r})$ need to be taken into account. 

The method proposed by Brueckner~\cite{brueckner1968} to get the potential (assumed to be central) between the nuclei at a distance $R$ is to compute the total energy of the system from the sum of the densities and subtracting the individual ground-state energies:
\begin{equation}
V(R) = E[\rho_1+\rho_2]-E[\rho_1]-E[\rho_2]
\end{equation}
where $\rho_1$ and $\rho_2$ are the HF ground-state one-body density matrices of the nuclei separated by a distance $R$.

It is important to note that, although the Pauli exclusion principle is properly accounted for between nucleons belonging to the same nucleus, it is neglected between nucleons of different nuclei. 
The problem comes from the fact that the total system is described by summing densities instead of building a properly antisymmetrised many-body wave-function. 

\begin{figure}
\centering
\includegraphics[width=6.5cm]{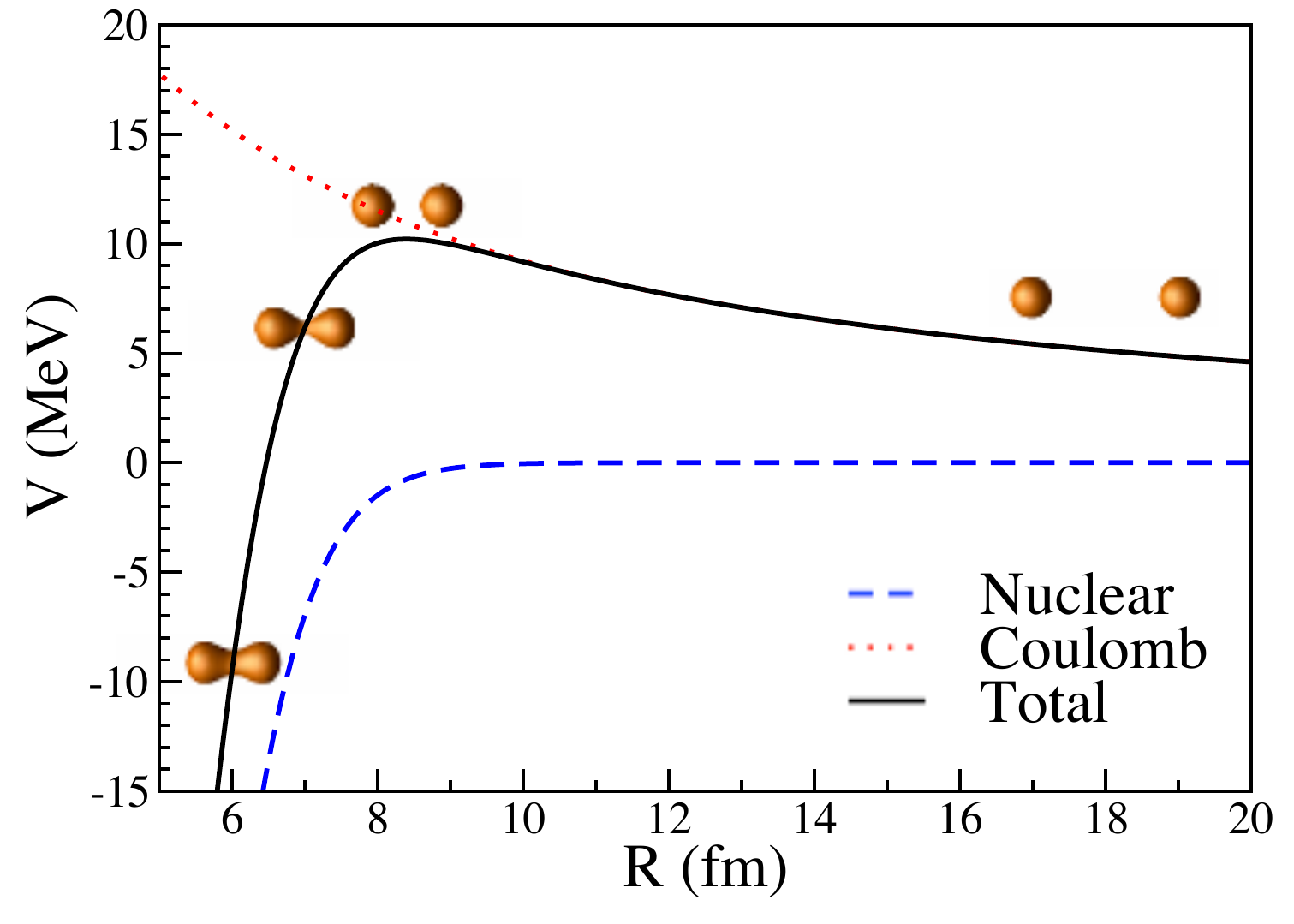}
\caption{Frozen Hartree-Fock nucleus-nucleus potential of $^{16}$O$+^{16}$O with the SLy4$d$ Skyrme parameterization.}
\label{fig:pot_O+O}       
\end{figure}

An example of such frozen Hartree-Fock (FHF) potential is shown in Fig.~\ref{fig:pot_O+O} for $^{16}$O$+^{16}$O. 
The nuclear part is computed from the SLy4$d$ parametrisation~\cite{kim1997} of the Skyrme functional and the Coulomb potential includes both the direct and exchange (using the Slater approximation) contributions.

The surfaces represent isodensities at half the saturation density $\rho_0/2=0.08$~fm$^{-3}$.
We see that, at the barrier, the nuclei are still relatively far from each other, indicating that the overlap between their densities is small. 
The Pauli exclusion principle is thus expected to play a minor role near the barrier in such light systems. 
Well inside the barrier, however, we see that the spatial overlap between the nuclei is more significant, and is then expected to lead to a ``Pauli repulsion''~\cite{fliessbach1971}.

\section{Pauli repulsion}
\label{sec:Pauli}

In the past, Pauli repulsion has been accounted for in the nucleus-nucleus bare potential with various methods. 
For instance, a direct antisymmetrisation of the overlapping wave-functions (e.g., with a Graam-Schmidt algorithm), has been considered  \cite{fliessbach1971,brink1975,zint1975}. 
The problem with this technique is that it can  potentially reduce the neck density  inducing too large Pauli repulsion~\cite{simenel2017}.
Another  traditional method  is to increase the kinetic  density $\tau(\mathbf{r})$ (e.g., via the Thomas-Fermi model)
\cite{brink1975,zint1975}.
This method, however, neglects  the effect of the Pauli exclusion principle on other terms of the functional, such as the spin-orbit term which has been shown to absorb a large part of the Pauli repulsion~\cite{simenel2017}.
Thus, the Pauli exclusion principle has a more complicated effect than just increasing the kinetic energy of the nucleons.

\subsection{The density-constrained frozen Hartree-Fock (DCFHF) method}

The density-constrained frozen Hartree-Fock (DCFHF) approach was proposed in Ref.~\cite{simenel2017}.
It is a new method to calculate microscopic bare nucleus-nucleus potentials  taking into account the Pauli exclusion principle exactly. 
It can be seen as the static counterpart of the well established density-constrained time-dependent Hartree-Fock (DCTDHF) method to compute instantaneous dynamical potentials in heavy-ion collisions \cite{umar2006b} (see section~\ref{sec:dynamics}). 

The DCFHF method is based on a similar variational principle as in Eq.~(\ref{eq:HF}) with additional constraints on the proton and neutron local densities,
\begin{equation}
\delta\,\langle\Phi|\left[\hat{H}-\int d^3r \,\,\lambda(\mathbf{r}) \rho(\mathbf{r})\right] |\Phi\rangle = 0,
\label{eq:DCFHF}
\end{equation}
where the distinction between proton and neutron densities has been omitted for clarity. 
The Lagrange parameters $\lambda(\mathbf{r})$ constrain the density at each point $\mathbf{r}$ to be the sum of the HF ground-sate densities of the nuclei at a distance $R$. 

The independent particle state $|\Phi\rangle$ thus describes the entire system and is fully antisymmetrised so that the Pauli exclusion principle is accounted for exactly.
It is used to compute the DCFHF potential via
\begin{equation}
V_{DCFHF}(R)= \langle\Phi|\hat{H}|\Phi\rangle -E[\rho_1]-E[\rho_2].
\label{eq:VDCFHF}
\end{equation}

\begin{figure}
\centering
\includegraphics[width=6.5cm]{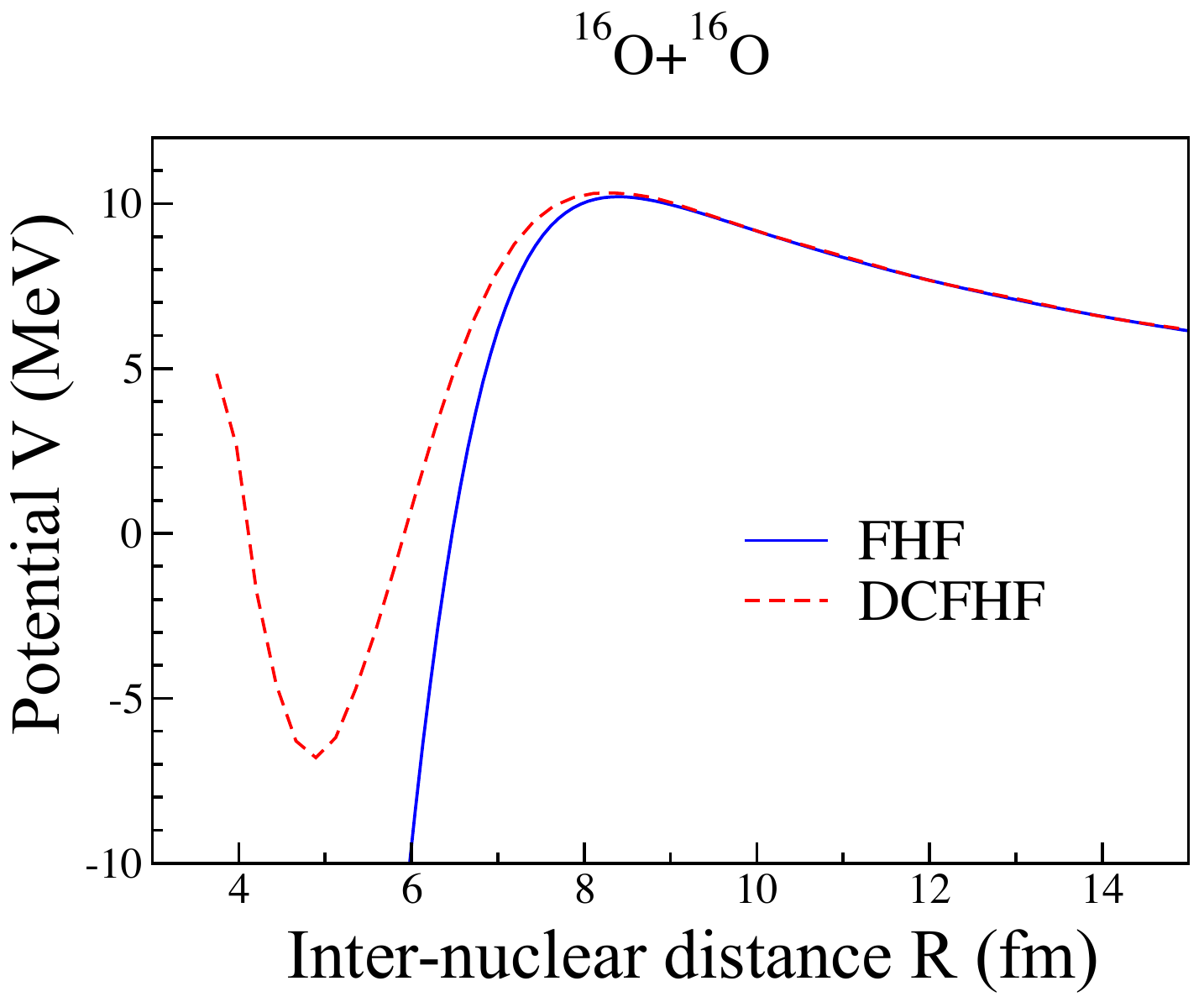}
\caption{Comparison of the FHF and DCFHF potentials in $^{16}$O$+^{16}$O with the SLy4d parameterization.}
\label{fig:O+O}       
\end{figure}

Figure~\ref{fig:O+O} shows the FHF and DCFHF potentials in $^{16}$O$+^{16}$O.
As expected, the inclusion of the Pauli exclusion principle has little effect near the barrier due to the small overlap between the nuclei.
However, a Pauli repulsion is observed inside the barrier, increasing its width.

\subsection{Charge product dependence}

It  is well known that the Coulomb barrier in light systems is obtained for small overlap between the nuclei (see Fig.~\ref{fig:pot_O+O}), while in heavy systems more overlap is required for the strong nuclear interaction to counterbalance the larger Coulomb repulsion. 
It is then interesting to investigate the charge product dependence of the Coulomb repulsion. 

\begin{figure}
\centering
\includegraphics[width=6.5cm]{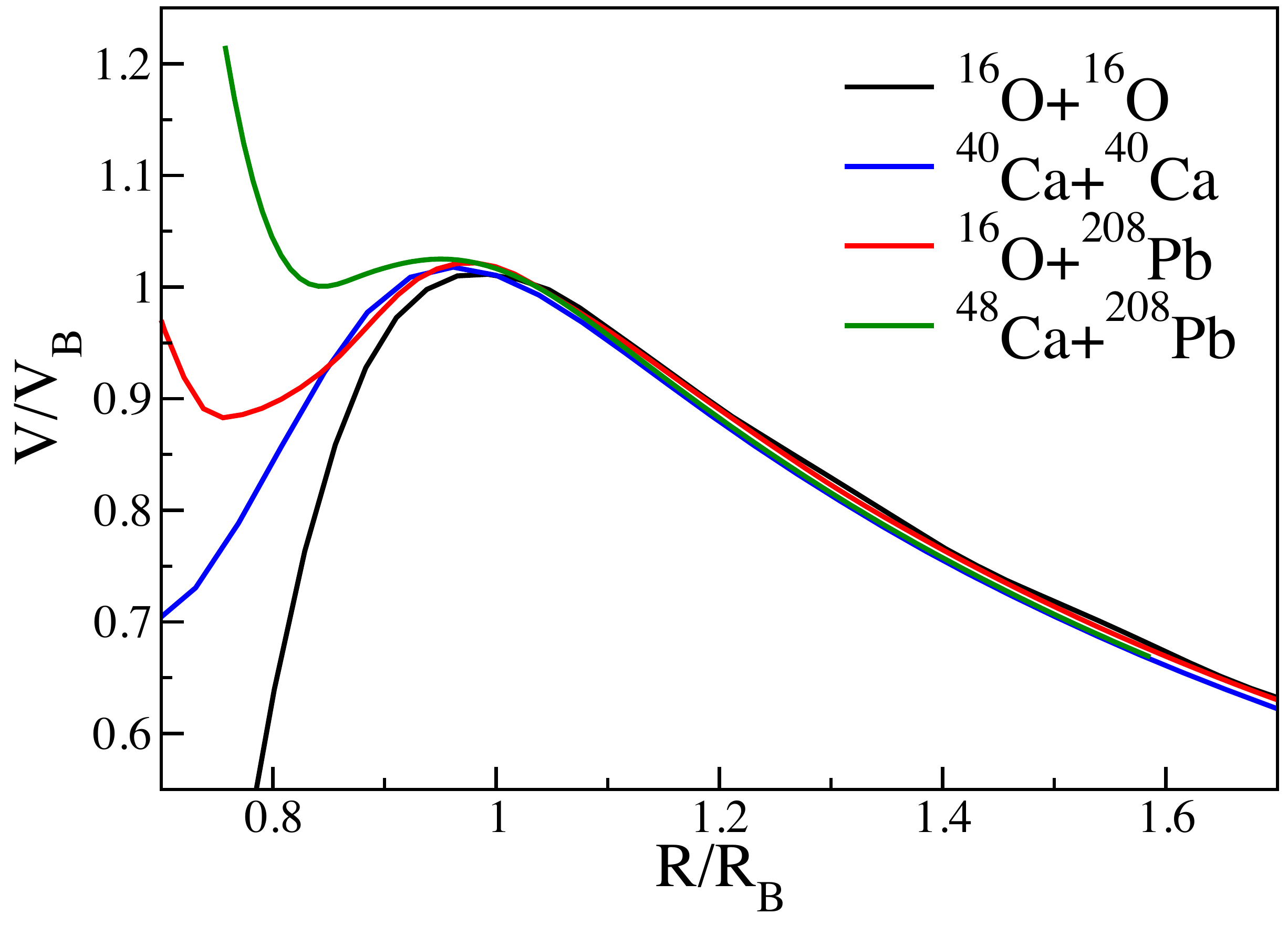}
\caption{DCFHF potentials in various systems. $R_B$ and $V_B$ denote the FHF barrier radius and height, respectively.}
\label{fig:V_Vb}       
\end{figure}

Figure~\ref{fig:V_Vb} shows a comparison of DCFHF potentials in various systems. 
We observe an important increase of the Pauli repulsion in the heavier systems. 
The pocket becomes shallower with increasing charge product $Z_1Z_2$ 
and almost disappears in $^{48}$Ca+$^{208}$Pb.
Note that the two-body  picture for such heavy systems is questionable. 
Indeed, the $^{48}$Ca+$^{208}$Pb case is extreme  as the DCFHF calculation predicts that fusion is impossible at $3\%$ below the barrier. 
In fact, a smooth transition toward an adiabatic potential for the compound system is expected \cite{ichikawa2009b} which would allow fusion to occur at lower energies. 

\section{Deep sub-barrier fusion hindrance}
\label{sec:subB}

Wether the pocket inside the barrier is physical or not, one expects an increase of the barrier width due to Pauli repulsion and a resulting reduction of the tunnelling probability. 
Pauli repulsion could then provide a possible contribution to the deep sub-barrier fusion hindrance observed experimentally~\cite{jiang2002,dasgupta2007,stefanini2010}. 

\subsection{Couplings to low-lying collective states}

In order to predict deep sub-barrier fusion cross-sections, it is important to take into account couplings to low-lying collective states. 
This is done traditionally within the coupled-channel (CC) framework (see Ref.~\cite{hagino2012} for a recent review). 
The CC approach starts with an ion-ion potential 
whose origin does not include any excitations of the nuclei, that is, 
the ``bare'' potential. In that sense, FHF and DCFHF potentials, being computed from ground-state densities, can in principle be used in CC calculations. 

However, in addition to the repulsion at short distance, the Pauli exclusion principle is expected to change the internal structure of the reactants and could then affect the coupling to low-lying collective states (see discussion in the supplemental material of Ref.~\cite{simenel2017}). In particular, it could induce a damping mechanism of collective vibration during the fusion process which requires extension of the standard coupled-channel method \cite{ichikawa2015b}.

\begin{figure}
\centering
\includegraphics[width=7cm]{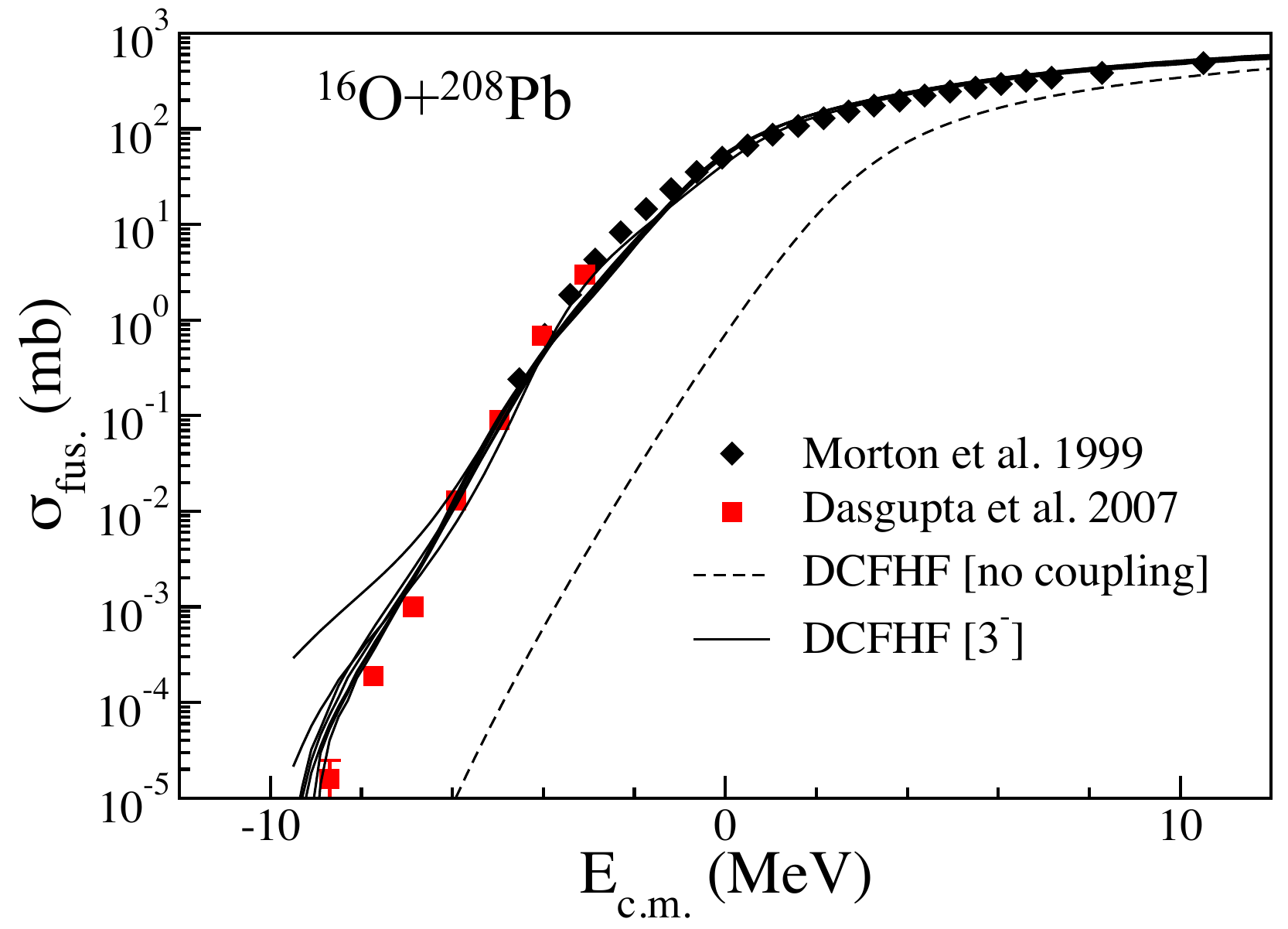}
\caption{Fusion cross-sections from \textsc{ccfull} with the DCFHF potential and couplings to the first collective $3^-$ state in target and projectile. The solid lines have been obtained with various choices of absorbing boundary conditions. Experimental data are from Refs.~\cite{morton1999,dasgupta2007}.}
\label{fig:OPb_CC}       
\end{figure}

Moreover, potentials with Pauli repulsion in medium-mass and heavy systems are often shallow (see Fig.~\ref{fig:V_Vb}), inducing numerical instabilities in CC calculations. 
This is illustrated in Fig.~\ref{fig:OPb_CC} where CC calculations of $^{16}$O$+^{208}$Pb fusion cross-sections down to deep sub-barrier energies have been performed with the \textsc{ccfull} code~\cite{hagino1999} using the DCFHF potential with different choices of absorbing boundary condition parameters.
Although the coupling to the first octupole phonon ($3^-$ state) in projectile and target produces a relatively good overall agreement with experimental data~\cite{morton1999,dasgupta2007}, we see that the behaviour at deep sub-barrier energies strongly depends on the choice of absorbing boundary conditions. 

Thus, to calculate fusion cross-sections at deep sub-barrier energies, 
in particular their logarithmic slopes, 
we use a 
simpler one-barrier penetration model and an overall shift of the fusion cross-sections 
in order to account for the overall effect of the couplings as suggested 
in Ref.~\cite{dasgupta2007}.

\begin{figure}
\centering
\includegraphics[width=7cm]{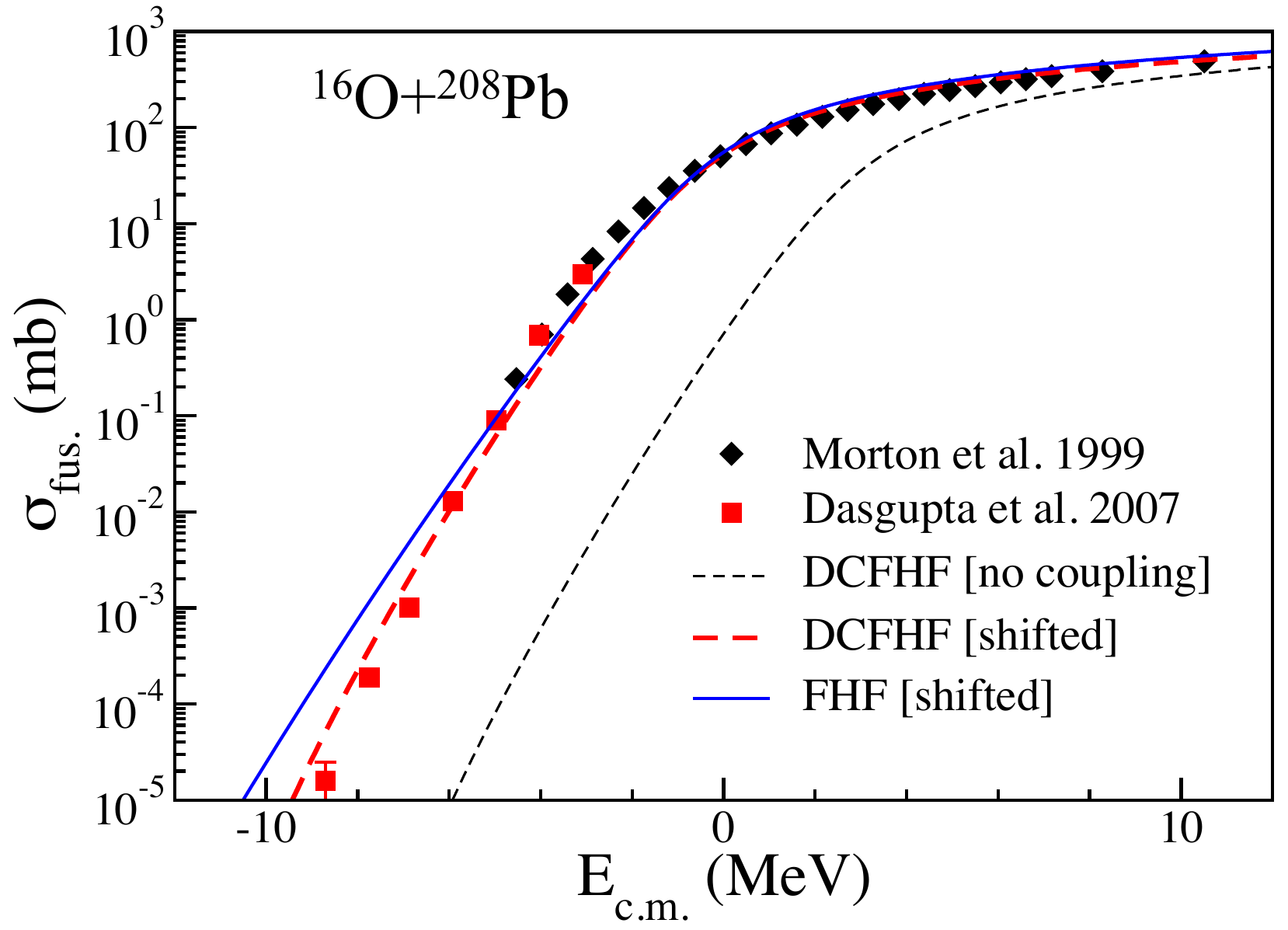}
\caption{Same as Fig.~\ref{fig:OPb_CC} with shifted cross sections to account for couplings.}
\label{fig:OPb_log}       
\end{figure}

\begin{figure}
\centering
\includegraphics[width=7cm]{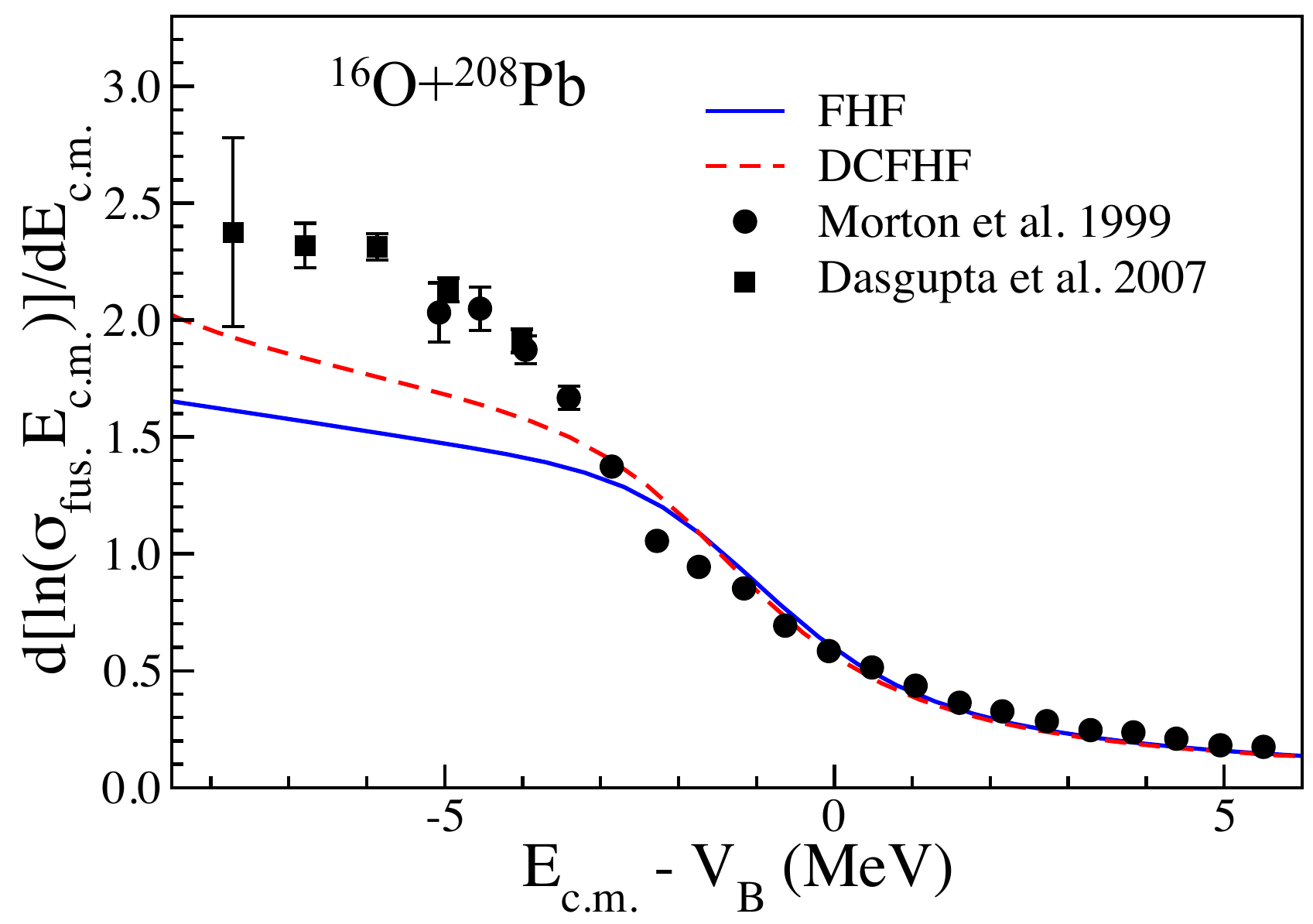}
\caption{Logarithmic slopes of the fusion cross-sections shown in Fig.~\ref{fig:OPb_log}.}
\label{fig:OPb_slope}       
\end{figure}

The results are shown in Figs.~\ref{fig:OPb_log} and~\ref{fig:OPb_slope}. 
We observe a better agreement with experimental data at deep sub-barrier energies for calculations with the DCFHF potential. 
In particular the logarithmic slope is improved. 
Nevertheless, discrepancies with experimental data remain, showing that the inclusion of the Pauli repulsion provides only part of the solution to the deep sub-barrier fusion hindrance problem. 

\subsection{Other hindrance mechanisms}

Let us pause to review briefly the other mechanisms invoked in the literature to explain deep sub-barrier fusion hindrance. 
\begin{itemize}
\item Mi\c{s}icu and Esbensen have argued that a pocket was formed inside the barrier due to the incompressibility of nuclear matter and the large density overlap between the nuclei~\cite{misicu2006}.
\item Ichikawa, Hagino and Iwamoto suggested that the hindrance could be due to a smooth transition from sudden to adiabatic potentials~\cite{ichikawa2009}. 
\item Dasgupta and collaborators invoked a possible decoherence mechanism reducing the effect of the couplings to low-lying collective states~\cite{dasgupta2007}.
\item The ANU group has investigated dissipation mechanisms induced by cluster transfer which could reduce the fusion probability~\cite{evers2011,rafferty2016}.
\item Ichikawa and Matsuyanagi have suggested the possibility of a reduction of the collectivity of vibrational modes on the way to fusion \cite{ichikawa2015b}.
\end{itemize}

The Pauli exclusion principle induces two additional possible fusion hindrance mechanisms.
The first one is due to the widening of the barrier from Pauli repulsion. 
The second one is the alteration of the couplings and collectivity of the low-lying states (an effect analogous to Ichikawa and Matsuyanagi's idea) due to the change of the internal structure induced by Pauli blocking. 

It is likely that more than one of the above effects contribute to the observed fusion hindrance at deep sub-barrier energies. 
However, the incompressibility is unlikely to play a strong role. 
Indeed, standard FHF calculations properly account for this effect as they are based on a Skyrme EDF fitted to reproduce the incompressibility of infinite nuclear matter. 
Yet, these calculations do not show any hindrance (see Figs.~\ref{fig:OPb_log} and~\ref{fig:OPb_slope}).
It is more likely that the repulsion potential introduced phenomenologically by Mi\c{s}icu and Esbensen in~\cite{misicu2006} is in fact simulating the effect of  the Pauli repulsion~\cite{simenel2017}.

\section{Dynamical nucleus-nucleus potential}
\label{sec:dynamics}

Apart from incompressibility, all the fusion hindrance mechanisms proposed above are in fact dynamical effects, involving time evolution in one way or another. 
We then naturally turn to a time-dependent generalisation of the DCFHF method to incorporate dynamical effects in the calculation of the potential. 

\subsection{The density-constrained time-dependent Hartree-Fock method}

The  density-constrained time-dependent Hartree-Fock (DCTDHF) method has been widely discussed and used in the literature~\cite{umar2006c,umar2006a,umar2012a,oberacker2013,jiang2014,umar2014a,umar2015a}.
To get a nucleus-nucleus DCTDHF potential, the starting point is a TDHF calculation of a collision above the fusion barrier. 

The TDHF equation is the time dependent generalisation of the variational principle~(\ref{eq:HF})
\begin{equation}
\delta\, \langle\Phi|(\hat{H}-i\partial_t)|\Phi\rangle=0.
\end{equation}
Modern TDHF codes \cite{kim1997,simenel2001,umar2006c,guo2012,maruhn2014,sekizawa2013,scamps2013a} have been used in a large number of fusion studies. See Refs.~\cite{washiyama2015,tohyama2016,bourgin2016,vophuoc2016,umar2016,shi2017} for recent applications to fusion. See also recent reviews~\cite{simenel2012,nakatsukasa2016}  and  contributions to this conference from A. S. Umar, L. Guo, K. Vo-Phuoc, K. Washiyama, G. Scamps, P. D. Stevenson, B. Schuetrumpf, K. Sekizawa and A. Bulgac for applications of TDHF and its extensions.

Figure~\ref{fig:tunnelling} shows an example of density evolution in $^{16}$O$+^{16}$O central collisions.
At above barrier energies, a set of densities leading to the compact fused system is obtained. 
The idea behind DCTDHF is then to use these densities as constraints in Eq.~(\ref{eq:DCFHF}) and to compute the potential from Eq.~(\ref{eq:VDCFHF}). 
The Pauli exclusion principle is included exactly as the TDHF density is obtained from a fully antisymmetrised state $|\Phi\rangle$ describing the entire system.
Note that, as the DCTDHF potential already incorporates dynamical effects, it is not a bare potential and should not be used in coupled-channels calculations. 
An effect of the couplings between relative motion and internal excitations is to induce an energy dependence of the potential~\cite{umar2014a}.

\subsection{(Isovector) transfer couplings}

In a recent work~\cite{godbey2017}, we used the following decomposition of the energy density
\begin{equation}
\mathcal{H}(\mathbf{r})=\frac{\hbar^2}{2m}\tau(\mathbf{r})+\mathcal{H}_0(\mathbf{r})+\mathcal{H}_1(\mathbf{r})+\mathcal{H}_C(\mathbf{r}),
\end{equation}
where $\mathcal{H}_0$,  $\mathcal{H}_1$ and  $\mathcal{H}_C$ are the isoscalar, isovector, and Coulomb  contributions, respectively, in order to separate the isoscalar ($v_0$) and isovector ($v_1$) contributions to the DCTDHF potential,
\begin{equation}
V(R)=v_0(R)+v_1(R)+V_C(R),
\end{equation}
where $V_C$ is the Coulomb potential. 

In the static case (e.g., FHF and DCFHF), the isovector potential $v_1$ vanishes. 
It is thus entirely induced by dynamical effects. 
For instance, systems with $N/Z$ asymmetries encounter a rapid charge equilibration (transfer of protons and neutrons in opposite directions)~\cite{simenel2012b} which has a strong impact on $v_1$~\cite{godbey2017}. 
This is illustrated in Fig.~\ref{fig:iso} for the $^{16}$O$+^{208}$Pb system, where a large reduction of the potential inside the barrier is induced by isovector transfer.
The resulting effect on the inner part of the potential is then opposite to the Pauli repulsion. 

\begin{figure}
\centering
\includegraphics[width=6cm]{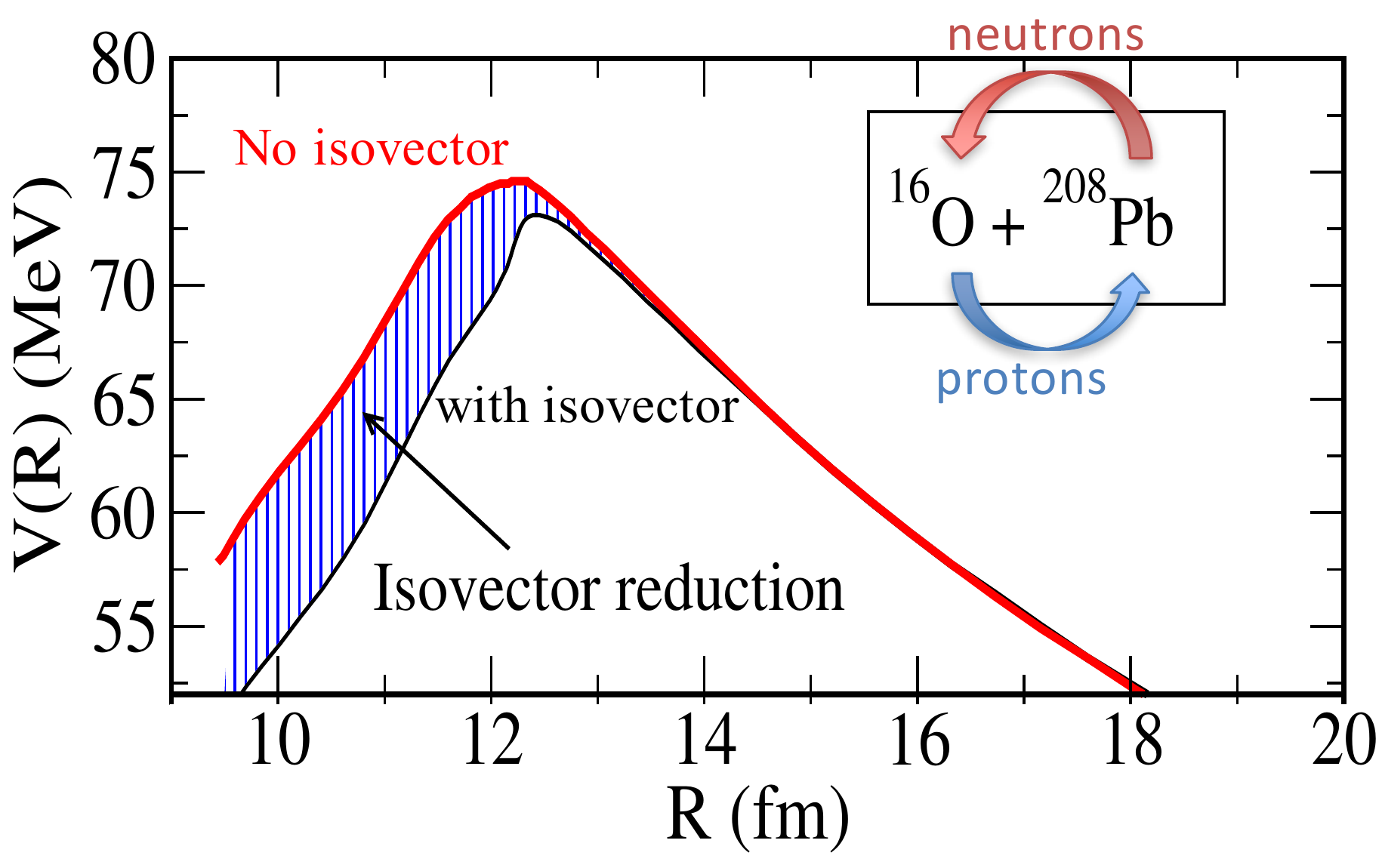}
\caption{DCTDHF potential in $^{16}$O$+^{208}$Pb with (black) and without (red) isovector contribution $v_1$ from a near-barrier TDHF evolution at $E_{c.m.}=75$~MeV. }
\label{fig:iso}       
\end{figure}

We then used this technique to investigate the impact of transfer on heavier systems where experimental signatures are not so clear~\cite{back2014,liang2016}.
The isovector reduction of the potential due to transfer channels depends naturally on the presence of positive $Q-$value transfer channels.
As shown in Fig.~\ref{fig:iso}, isovector reduction is observed in  $^{40}$Ca$+^{132}$Sn which has several positive $Q-$value transfer channels, but not in  $^{48}$Ca$+^{132}$Sn which has no  positive $Q-$value transfer channel.
This is also confirmed by the proton number distributions in the heavy fragments obtained just below the barrier from a particle number projection technique~\cite{simenel2010} which show almost no proton transfer in  $^{48}$Ca$+^{132}$Sn (left panels in Fig.~\ref{fig:iso}; see also Ref.~\cite{vophuoc2016} and K. Vo-Phuoc's contribution).  A similar effect (not shown in Fig.~\ref{fig:iso}) is observed for  neutron transfer.  
\begin{figure}
\centering
\includegraphics[width=8cm]{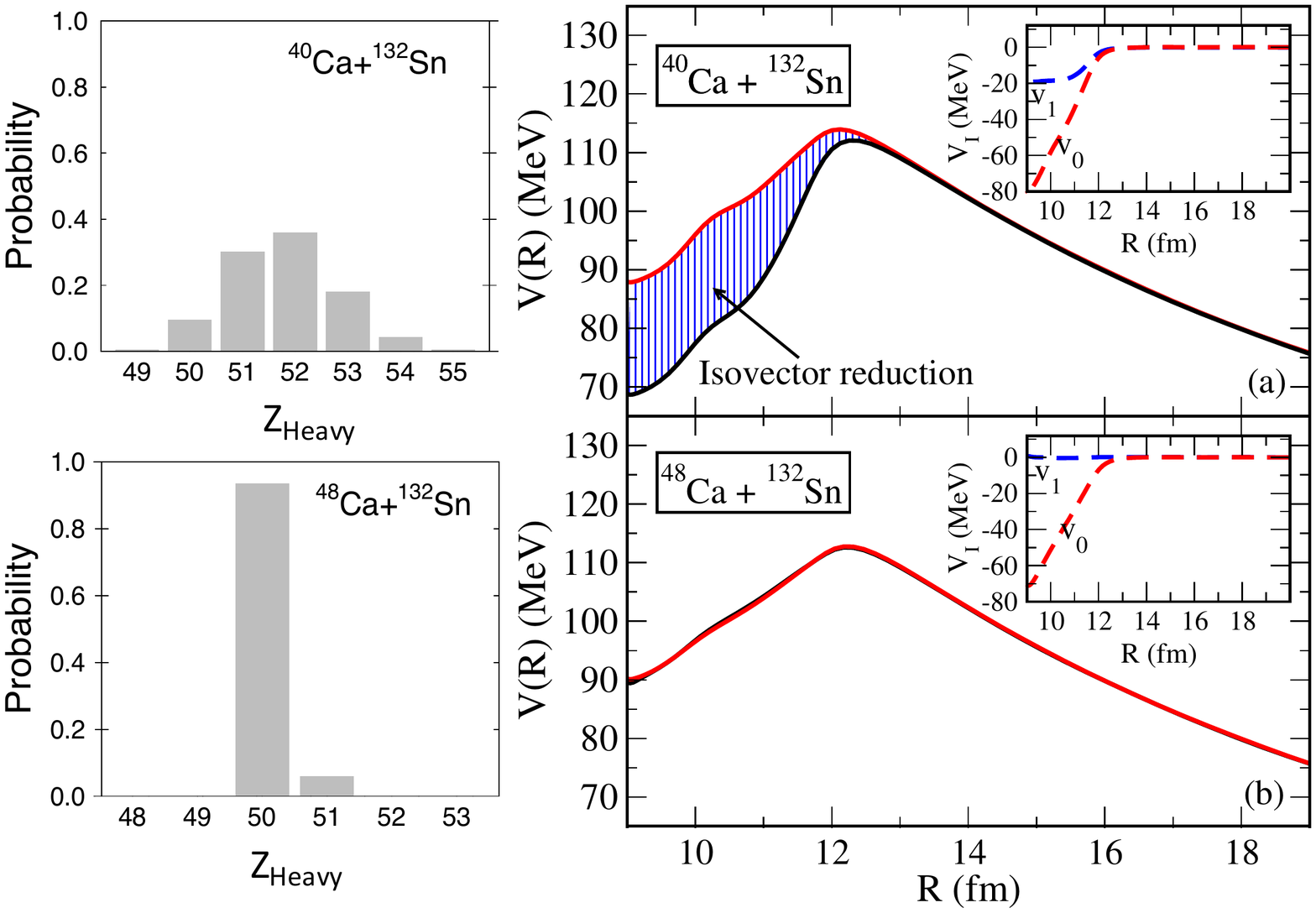}
\caption{(Right) Same as Fig.~\ref{fig:iso} for $^{40,48}$Ca$+^{132}$Sn reactions at $E_{c.m.}=120$~MeV. (Left) final proton number probability distributions in the heavy fragment just below the barrier.}
\label{fig:iso}       
\end{figure}

The above analysis provides an explanation for the fact that fusion in the $^{48}$Ca$+^{132}$Sn system is relatively well explained by standard  CC calculations neglecting transfer, whereas they underpredict the fusion cross sections in $^{40}$Ca$+^{132}$Sn~\cite{kolata2012}. Indeed, transfer is expected to increase sub-barrier fusion in the latter system thanks to a narrowing of the barrier (see Fig.~\ref{fig:iso}-a).
As DCTDHF potentials account both for transfer effects and Pauli repulsion, the calculated sub-barrier fusion cross-sections from such potentials are in relatively good agreement with experiment~\cite{oberacker2013}, despite the fact that these calculations have no adjustable parameters.

\acknowledgement
 This work has been supported by the
Australian Research Council Grants Nos. FT120100760,
and by the U.S. Department of Energy under grant Nos.
DE-SC0013847.
 \bibliography{VU_bibtex_master}

\end{document}